\documentclass[conference]{IEEEtran}
\IEEEoverridecommandlockouts
\usepackage{cite}
\usepackage{amsmath,amssymb,amsfonts}
\usepackage{algorithmic}
\usepackage{graphicx}
\usepackage{textcomp}
\usepackage{xcolor}
\usepackage[a4paper, total={184mm,239mm}]{geometry}
\usepackage{xspace}
\usepackage{etoolbox}
\usepackage[textsize=scriptsize,textwidth=1.1cm]{todonotes}
\usepackage{lipsum}
\usepackage{listings}
\usepackage{url}
\usepackage{comment}

\definecolorseries{test}{rgb}{grad}[rgb]{.95,.55,.55}{11,11,17}
\resetcolorseries[10]{test}
\newcommand{\addtodoeditor}[1]{%
    \colorlet{#1}{test!!+!50}
    \expandafter\newcommand\csname#1\endcsname [1]{%
        \todo[color=#1,size=\tiny]{\sffamily\textbf{\uppercase{#1}:}
    ##1}\xspace%
    }
    \expandafter\newcommand\csname#1i\endcsname [1]{%
        \todo[inline, color=#1]{\sffamily\textbf{\uppercase{#1}:} ##1}\xspace%
    }
}

\addtodoeditor{ab}
\addtodoeditor{df}
\addtodoeditor{am}

\newcommand{\cvm}[1]{{\textit{cVM#1}}\xspace}
\newcommand{\iperf}{{\emph{iperf3}}\xspace}
\newcommand{\musl}{\textit{musl libc}\xspace}

\def\BibTeX{{\rm B\kern-.05em{\sc i\kern-.025em b}\kern-.08em
    T\kern-.1667em\lower.7ex\hbox{E}\kern-.125emX}}
\begin{document}
\iftrue
\title{Enabling Security on the Edge: A CHERI Compartmentalized Network Stack
\thanks{A. Bastoni and A. Zuepke were supported by the Chair for Cyber-Physical Systems in Production Engineering at TUM and the Alexander von Humboldt Foundation. This work has also been supported by Secure Systems Research Center, Technology Innovation Institute (TII). The work was also supported by the European Union under the NextGenerationEU Programme within the Plan ``PNRR - Missione 4 ``Istruzione e Ricerca'' - Componente C2 Investimento 1.1 ``Fondo per il Programma Nazionale di Ricerca e Progetti di Rilevante Interesse Nazionale (PRIN)'' by the Italian Ministry of University and Research (MUR)''. Project Title: ``Simplifying Predictable and energy-efficient Acceleration from Cloud to Edge (SPACE)'', Project code: E53D23007800006. MUR D.D. financing decree n. 959, 30th June 2023.}
}
\author{%
\IEEEauthorblockN{%
Donato Ferraro\IEEEauthorrefmark{1}\IEEEauthorrefmark{2}, %
Andrea Bastoni\IEEEauthorrefmark{2}\IEEEauthorrefmark{3}, %
Alexander Zuepke\IEEEauthorrefmark{2}\IEEEauthorrefmark{3}, %
Andrea Marongiu\IEEEauthorrefmark{1}
}

\IEEEauthorblockA{%
University of Modena and Reggio Emilia\IEEEauthorrefmark{1}, %
Minerva Systems\IEEEauthorrefmark{2}, %
Technical University of Munich\IEEEauthorrefmark{3}
}
}
\else
\title{Enabling Security on the Edge: A CHERI Compartmentalized Network Stack
\thanks{}
}
\author{%
\IEEEauthorblockN{%
First Author, %
Second Author, %
Third Author, %
Fourth Author
}
}
\fi


\maketitle

\begin{abstract}
The widespread deployment of embedded systems in critical infrastructures, interconnected edge devices like autonomous drones, and smart industrial systems requires robust security measures.
Compromised systems increase the risks of operational failures, data breaches, and---in safety-critical environments---potential physical harm to people.
Despite these risks, current security measures are often insufficient to fully address the attack surfaces of embedded devices.
CHERI provides strong security from the hardware level by enabling fine-grained compartmentalization and memory protection, which can reduce the attack surface and improve the reliability of such devices.
In this work, we explore the potential of CHERI to compartmentalize one of the most critical and targeted components of interconnected systems: their network stack.
Our case study examines the trade-offs of isolating applications, TCP/IP libraries, and network drivers on a CheriBSD system deployed on the Arm Morello platform.
Our results suggest that CHERI has the potential to enhance security while maintaining performance in embedded-like environments.
\end{abstract}


\section{Introduction}

Nowadays, interconnected, powerful embedded devices are widely used ``at the edge,'' such as in drones (UAVs), autonomous vehicles, smart industrial plants, and in many components of critical infrastructures.
Although these devices are relatively powerful, compared to high-performance data center systems, edge devices are more resource-constrained in terms of processing power and memory. They also typically operate in critical environments with real-time requirements.

To improve performance, reduce size, and save space, many of these devices lack security features such as Memory Protection Units (MPUs) and Memory Management Units (MMUs).
As a result, all applications typically run within a single address space.
For example, the \mbox{NuttX}/\mbox{PX4} framework~\cite{PX4, PX4-paper}, widely used in both autonomous and remote-controlled drones, runs all software components---network stacks, drivers, and applications (e.g., actuation)---without isolation.

This lack of isolation increases the risk that a vulnerability in one component could compromise the entire system. For instance, a buffer overflow in the network stack could allow an attacker to take full control of a drone, potentially causing physical damage (e.g., making the drone crash) or leaking sensitive data, such as video surveillance footage.
As embedded systems are increasingly used in sensitive and highly connected environments, the need to ensure their effective security is becoming more critical~\cite{fut-app-iot}.
Unfortunately, even when MPUs and MMUs are available, their coarse-grained isolation can lead to performance degradation and reduced predictability~\cite{MxU,COTS}. This issue also affects systems that use lightweight isolation among threads to minimize the overhead of task-switches.


Capability Hardware Enhanced RISC Instructions (CHERI)~\cite{CHERIISA} represents an advancement in addressing the aforementioned security challenges. CHERI extends conventional CPU architectures with a \textit{capability}-based security model, providing fine-grained memory protection at byte level, by associating each memory reference with a \textit{capability}---a protected token that specifies bounds and access rights to it. This mechanism allows systems to compartmentalize software components efficiently and prevents them from accessing memory regions outside their designated bounds. Unlike traditional memory isolation techniques that rely on MPUs or MMUs, CHERI enforces security at the hardware level with potentially lower overhead, making it a promising candidate for embedded systems where resources are constrained, and enhanced memory protection is critical.


However, CHERI is still an emerging area of research, and aside from FPGA deployments, there is limited availability of CHERI-enabled hardware. The CHERI RISC-V Sonata development platform~\cite{Sonata} was only unveiled in 2024, and the Arm Morello~\cite{Morello} platform is available on request only.
The Arm Morello platform is currently the only Arm-based prototype supporting CHERI. Although it is more powerful than typical microcontrollers and small embedded systems, it provides a valuable real-world testbed for exploring the potential and the trade-offs of CHERI, even for embedded-like devices. For example, the platform has been used in the AutoCHERI project~\cite{AutoCHERI} to evaluate CHERI’s feasibility in enhancing security for autonomous vehicles.


Given the central role of the network stack in interconnected edge devices, it is unsurprising that network components are major targets of security attacks~\cite{emb-sys-sec,sec-iiot,trends-rahimi}, with memory-related issues being one of the main intrusion vectors.
For instance, recent vulnerabilities like CVE-2023-52370~\cite{CVE-2023-52370} and CVE-2023-6951~\cite{CVE-2023-6951} exploit buffer overflows in the network stack, while CVE-2024-38951~\cite{CVE-2024-38951} leverages unchecked buffer limits to mount a Denial-of-Service (DoS) attack on the MAVLink\cite{mavlink} protocol of \mbox{PX4}.
Compartmentalizing the elements of the network stack can prevent vulnerabilities in one component from compromising the entire system.

The network stack is typically composed by (i)~an Ethernet driver, responsible for managing the physical Network Interface Card (NIC); (ii)~a TCP/IP protocol library; and, (iii)~applications.
In this work, we analyze how to compartmentalize these components by leveraging CHERI fine-grained protection.
Specifically, we aim to evaluate the trade-offs and performance of encapsulating the different components into CHERI compartments. Due to the mentioned hardware limitations, to obtain representative measurements, we conduct our evaluation on the Arm Morello platform using the CheriBSD operating system (OS)~\cite{CheriBSD}.
We assess the behavior of single address space systems using the Data Plane Development Kit (DPDK)~\cite{dpdk}, a framework that allows direct access to network hardware from user space, bypassing the networking stack of the operating system. Both DPDK and the F-Stack TCP/IP library~\cite{fstack} were ported to CHERI Morello.
A modified version of the CAP-VM Intravisor~\cite{CAPVM} was used to experiment with different isolation configurations.
This work highlights two key points: (i)~CHERI is a promising architecture for enhancing security in low-end embedded systems, as well as in scenarios where thread isolation is required to minimize overhead, and (ii)~the overhead introduced by this architecture is minimal.


\section{Background}\label{sec:background}

\subsection{CHERI}
CHERI~\cite{CHERIISA} is a technology that enhances Instruction Set Architectures (ISAs)---e.g., RISC-V, Arm---with \textit{capability}-based primitives, improving software security and preventing vulnerabilities. 
CHERI introduces two key concepts.

\smallskip
\noindent
\textbf{Fine-grained code protection using capabilities.}
CHERI replaces traditional pointers with ``capability pointers''. These pointers include metadata such as bounds, access permissions, and integrity tag---which carry information on the validity of a \textit{capability}.
These metadata are stored within the pointer, doubling its size. Hence, the size of the \textit{capability} is twice the size of the native platform register size---i.e., a 64-bit platform has \textit{capabilities} of 128-bit.
The use of \textit{capabilities} restricts memory access and permissions at a granular level, ensuring that software can only access what it is authorized to. \textit{Capabilities} can only derive from valid \textit{capabilities} (valid provenance) and new \textit{capabilities} are produced with less or equal privilege as their parent \textit{capability} (monotonicity), i.e., read \textit{capability} cannot form a read/write \textit{capability}.
CHERI provides two different types of compilation: \textit{hybrid} mode, where the programmer specifies which pointers are \textit{capabilities} using keywords in the C code, and \textit{pure} mode, where every pointer is treated as a \textit{capability}. This dual-mode approach allows legacy code to run unmodified while gradually porting software to a fully \textit{capability}-based protection system.

\smallskip
\noindent
\textbf{Secure software compartmentalization.}
\textit{Capabilities} in CHERI are used to enforce isolation within the same address space. Different components of a program can be isolated from each other, with each \emph{compartment} having its own set of permissions and memory bounds, making it harder for malicious code to affect other parts of the system.

In \textit{hybrid} architectures, compartmentalization is preserved by new special \textit{capability} registers: the Default Data Capability (\textit{DDC}), which defines the boundaries of the compartment, and the Program Counter Capability (\textit{PCC}), which defines the \textit{capability} of the program counter. This means that, if an application tries to execute a load or store instruction that goes outside its DDC, the architecture will generate a signal decoded as \textit{Capability Out-of-Bounds exception}.

CHERI \emph{sealing} implements robust compartmentalization, where \textit{capabilities} are restricted to specific tasks or regions of code. Through sealing, a \textit{capability} can be locked (or sealed) to a specific code or data and can only be unlocked through a special mechanism.


\subsection{CAP-VMs}

CAP-VMs~\cite{CAPVM}, is an example of a dynamic approach to compartmentalization with CHERI. CAP-VMs provide VM-like abstractions to isolate system applications---in a single-address-space environments---by leveraging CHERI memory \textit{capabilities}.
The CAP-VM approach comprises three main elements.
(i)~A host OS Kernel: a CHERI-aware kernel that provides the primitives for synchronization, execution contexts, and I/O operations. Similarly to CAP-VM~\cite{CAPVM}, we used CheriBSD.
(ii)~The \emph{Intravisor}: a process responsible to manage configuration, isolation, and lifecycle of \textit{capability}-VMs (cVMs). It distributes memory \textit{capabilities} to cVMs and has access to all cVM memory regions.
(iii)~\emph{capability-VM} (cVM): an isolated application component (\textit{hybrid} or \textit{pure} mode) that run as a thread of the Intravisor.

The Intravisor is in charge of the configuration of system \textit{capabilities} among the various cVMs, i.e., of the definition of the isolated memory ranges.
For cVMs, the Intravisor also acts as a proxy between the cVM \textit{capability} world and CheriBSD. cVMs do not have direct access to the host OS syscalls, but must use instead a \textit{trampoline} proxy table provided by the Intravisor that correctly handles the \textit{capabilities} and mediates the access to the OS.

In this work, we leverage the minimal trusted computing base (TCB) of the Intravisor, which makes it practical for integration into embedded systems.

\subsection{DPDK}


The \textit{Data Plane Development Kit}~\cite{dpdk} is an Open Source set of libraries and drivers designed to accelerate packet processing in high-performance networking applications.
DPDK bypasses the traditional kernel network stack and interacts directly with network hardware. This leads to better performance, but less isolation with the application. DPDK also operates in polling mode to reduce the latency caused by interrupt-triggered context switches. To detach the NIC from the kernel-space, DPDK uses a kernel-module.

Using DPDK, the management of network devices can be encapsulated within a cVM, thus isolating network operations from other system components and leveraging CHERI’s \textit{capability}-based security to ensure compartmentalization.

DPDK does not provide any TCP/IP protocol implementation, but only focuses on packet I/O and network hardware acceleration. Applications using DPDK must either implement their own protocol stack or integrate an existing user-space networking stack like the Open-Source F-Stack library~\cite{fstack}.
In this paper, we ported F-Stack on our CHERI architecture and integrated it in our DPDK cVMs.
Compared to other TCP/IP libraries built on top of DPDK, the existing integration of F-Stack with the FreeBSD network stack facilitated a more efficient porting process to CheriBSD.

We selected \iperf~\cite{iperf3} as an application for the evaluation of our compartmentalized network.
\iperf allows to define a server-client connection to measure the maximum bandwidth achievable.


\section{Network Stack Compartmentalization Options}\label{sec:architecture}

Different alternatives exist for the compartmentalization of a network stack using CHERI~\cite{cheriot-netstack, compartOS, cherirtos}.
Each alternative entails trade-offs regarding its intrusiveness (i.e., the amount of changes required), overheads, and potential scalability.
To this end, this work focuses on two possible system designs and compares them with a baseline scenario that does not use CHERI.

While we target resource-constrained systems that lack additional memory protection mechanisms, the trade-offs presented are also applicable to user-space device management with strong inter-thread isolation. This design is particularly efficient when low overhead is essential, leveraging multi-threading solutions to achieve this balance.



\begin{figure}
    \centering
    \includegraphics[width=\linewidth]{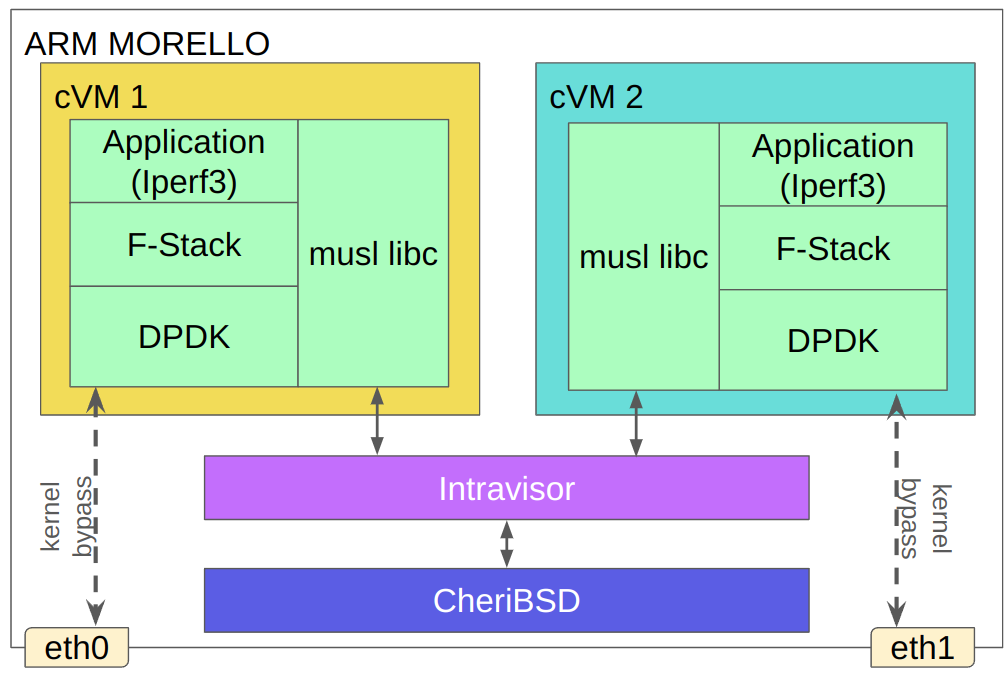}
    \caption{Scenario~1: Replication of the entire stack into two different \cvm{s}. The arrow shows where the \textit{trampoline} function is executed.}
    \label{fig:Scenario1}
\end{figure}

An overview of our CHERI-enabled designs---namely, Scenario~1 and Scenario~2---is presented in Figure~\ref{fig:Scenario1} and Figure~\ref{fig:Scenario2.1}.
Additionally, we define a \textit{Baseline} scenario for comparison. 
This serves as a reference point for evaluating the two CHERI-enabled scenarios.

We run our experiments on the Arm Morello on top of the CheriBSD host OS. For the two CHERI-enabled scenarios, the Intravisor is responsible for the configuration of the compartments and mediates the syscalls between compartments and CheriBSD.
Since the Morello built-in Ethernet device is not supported by DPDK, we chose a PCI card \textit{Intel 82576 Gigabit Network Connection} with two Ethernet ports.
%

\subsection{Scenarios}

\smallskip
\noindent
\textbf{Baseline.}
\textit{Baseline} consists of a non-CHERI full-network stack. In this configuration, DPDK, F-Stack, and \iperf run on two different processes (comparison with Scenario~1), and as a single process (Scenario~2). This setup represents a system that uses the MMU to isolate the two processes, which limits the ability to achieve fine-grained isolation. 

\smallskip
\noindent
\textbf{Scenario~1.}
In Scenario~1, each compartment (\cvm{1}, \cvm{2}) contains one network application (\iperf), the F-Stack TCP/IP library, and the DPDK user-space network layer.
The components run in \textit{hybrid} mode and are linked against a modified \musl library that provides the \textit{trampoline} wrappers towards the Intravisor.
Each \cvm{} works as independent network processing unit, with its own dedicated stack and Ethernet interface (``eth0'' and ``eth1'').
This configuration provides a complete compartmentalization approach, ensuring that any security breach does not affect the other \cvm{s}. Moreover, it also serves as a baseline for comparing performance and isolation. The \textit{trampoline} function is only executed when \musl syscalls are executed.

\smallskip
\noindent
\textbf{Scenario~2.}
\begin{figure}
    \centering
    \includegraphics[width=\linewidth]{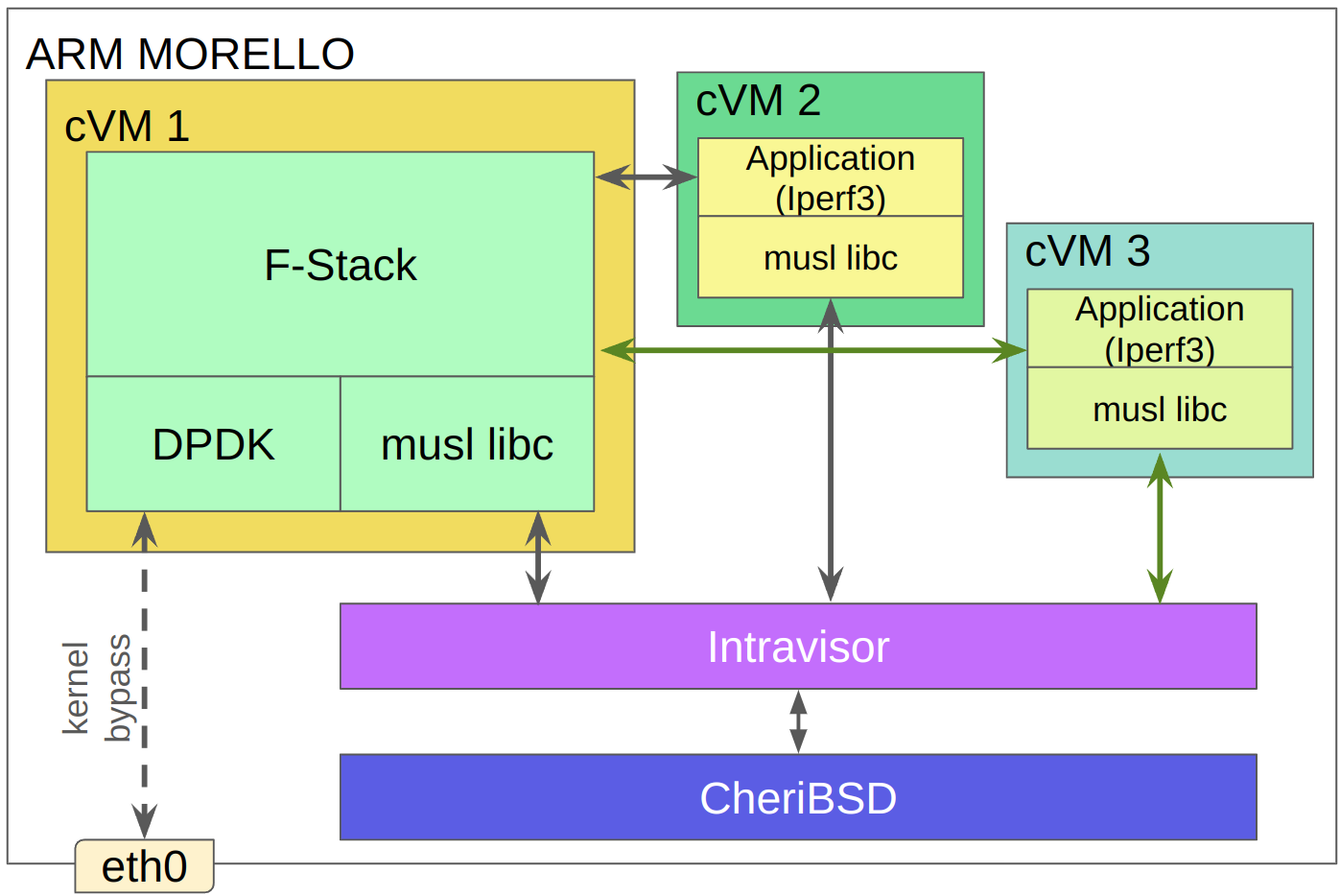}
    \caption{Scenario~2: Isolation of two distinct applications \cvm{s} from a F-Stack and DPDK cVM in the contented variant.}
    \label{fig:Scenario2.1}
\end{figure}
This scenario separates the application (\iperf) from F-Stack and DPDK, thus allowing for potential space optimization in constrained environments.
Here, \cvm{1} runs the TCP/IP library and DPDK stack, while \cvm{2} (and \cvm{3}) execute (separate instances of) the application.
All compartments link against the modified \musl library to interact with the Intravisor.
\cvm{1} is responsible to read/write from/to the Ethernet driver queues and sort to the right file descriptor, when requested. This scenario requires a \textit{mutex} to coordinate the execution of the F-Stack API functions and the main-loop execution, which creates a potential contention issue for this shared resource.
Therefore, we have evaluated this scenario in two configurations (see Section~\ref{sec:evaluation}) with one application (uncontented) or two compartmentalized applications (contented) respectively.
Figure~\ref{fig:Scenario2.1} depicts the contented configuration, i.e., with two compartmentalized applications.

\subsection{Implementation}\label{sec:implementation}


\smallskip
\noindent
\textbf{Intravisor.}
In CAP-VMs~\cite{CAPVM}, \textit{hybrid} \cvm{s} encapsulating applications
use a combination of \musl~\cite{musl} and Linux Kernel Library (LKL)~\cite{lkl}. LKL turns the Linux kernel into a library, providing kernel services---such as file systems and networking---to be run in user-space, making the resulting application highly portable to various environments---e.g., virtualized and bare-metal. However, this may increase the attack surface, and introduces overhead by emulating kernel syscalls in user-space.
Conversely, our DPDK and F-Stack are designed to fully execute in user-space and interact with the kernel only at boot time. They can therefore be streamlined within \cvm{s}, removing the additional LKL layer.
We directly connected \musl in the Intravisor substituting supervisor call instructions (\texttt{svc}) with dedicated \textit{trampoline} functions.
Specifically, a \textit{trampoline}
passes through the syscall ID and arguments, stores register states.
It also loads the correct PCC and DDC, and use them to jump into the cVM/Intravisor using CHERI specific instruction (e.g., \texttt{blrs} for the Arm Morello).
Avoiding LKL helps reducing overhead as well as the overall footprint and attack surface, making the \cvm{s} potentially more secure.
Similarly to FreeBSD, CheriBSD native libc is not fully compatible with \musl. 
We adapted 
the Intravisor proxy function to properly translate \musl calls into CheriBSD libc equivalents. For instance, \musl uses \textit{futex} for thread synchronization, while CheriBSD uses \textit{umtx}~\cite{umtx}.


\smallskip
\noindent
\textbf{DPDK.}
DPDK was ported to the CHERI Morello using \textit{hybrid} mode~\cite{dpdk-morello}. However, the existing support is very limited and does not initialize nor use any NIC interface.
We implemented the module that detaches the NIC from kernel-space and attaches it to user-space, ensuring that the memory allocations it requests are performed with the correct permission flags. Minor adjustments were required to enable execution using \musl.

\smallskip
\noindent
\textbf{F-Stack.}
F-Stack leverages the FreeBSD network stack, but, prior to this work, no available implementations of F-Stack for CHERI existed.
The first step in our implementation involved porting F-Stack to work with CheriBSD. This required adaptations on both OS and F-Stack sides.
CheriBSD was modified to accommodate the specific requirements of F-Stack, by streamlining the use of the network stack.
DPDK and F-Stack operate in polling mode, meaning that the application must explicitly request to read/write from/to the device. After an initialization phase, a main-loop is executed, with the key tasks being: (i)~process the ring buffers of the DPDK Ethernet driver; and, (ii) execute a user-defined function where calls to F-Stack API functions can be made.
F-Stack also required adaptations to work with \textit{capabilities}. 
We extended its data structures to use \textit{capabilities} and we have extended its Application Programming Interface (API).\footnote{The code for both DPDK and F-Stack extensions is available at \url{https://github.com/donato-ferraro/dpdk-morello-capvms} and \url{https://github.com/donato-ferraro/fstack-cheri}}
Applications can integrate with F-Stack through its API, which closely resembles the BSD socket API.
For instance, F-Stack exposes \texttt{ff\_socket()} and \texttt{ff\_write()} functions that are the equivalent of
\texttt{socket()} and \texttt{write()}. This allows developers to adapt existing network applications for F-Stack with minimal changes.
To work with \textit{capabilities}, the signatures of such APIs has to be modified.
For instance, the signature of the \texttt{ff\_write()} has been modified as follow:
\begin{lstlisting}[linewidth=\columnwidth,breaklines=true]
 -   ssize_t ff_write(int fd, const void * buf, size_t nbytes); 
 +   ssize_t ff_write(int fd, const void * __capability buf, size_t nbytes);
\end{lstlisting}

The total amount of modified lines of code (LoC) to achieve our objectives is shown in Table~\ref{tab:loc}.
Since the FreeBSD version originally supported by F-Stack is v13.0, while we used CheriBSD v$23.11$ based on FreeBSD v14.0, the modifications in Table~\ref{tab:loc} include not only a transition to a CHERI-enabled library, but also the adjustments needed to adapt the native library to the version of CheriBSD. Overall, this is in line with the usually expected modifications to port a library~\cite{HWenabled}.

\begin{table}[h]
    \centering
    \caption{Number of line of codes added/modified}
    \begin{tabular}{|c|c|c|}
    \hline 
        Library & LoC & global amount in percentage \\
        \hline
        F-Stack & $152$ & $0.99\%$ \\
    \hline
    \end{tabular}
    \label{tab:loc}
\end{table}

\smallskip
\noindent
\textbf{Application.}
To work with our setup, we initially ported \iperf to work with the F-Stack API. Next, we replaced the \texttt{select} function, with the \texttt{epoll} mechanism, which adapts better to F-Stack.
For Scenario~2, we also implemented the wrapper functions to the API of F-Stack to do the cross-compartment jump between the running application and the \cvm{1}.

\section{Evaluation}\label{sec:evaluation}

In this section, we evaluate the performance and security benefits of the system architecture and isolation configurations implemented on the Morello platform.
In addition to verifying the security achieved through CHERI compartmentalization, the evaluation focuses on two main aspects:
performance in terms of TCP bandwidth and execution time cost of a representative function, the \texttt{ff\_write()}.
To measure these values, we use \texttt{clock\_gettime()} with CLOCK\_MONOTONIC\_RAW.
The \texttt{ff\_write()} execution time serves as a meaningful performance indicator because this function has a direct impact on the overall system efficiency.
Note that, since DPDK and F-Stack operate in polling mode to minimize latency, CPU utilization is not a meaningful indicator.



\begin{figure}
    \centering
    \includegraphics[width=1\linewidth]{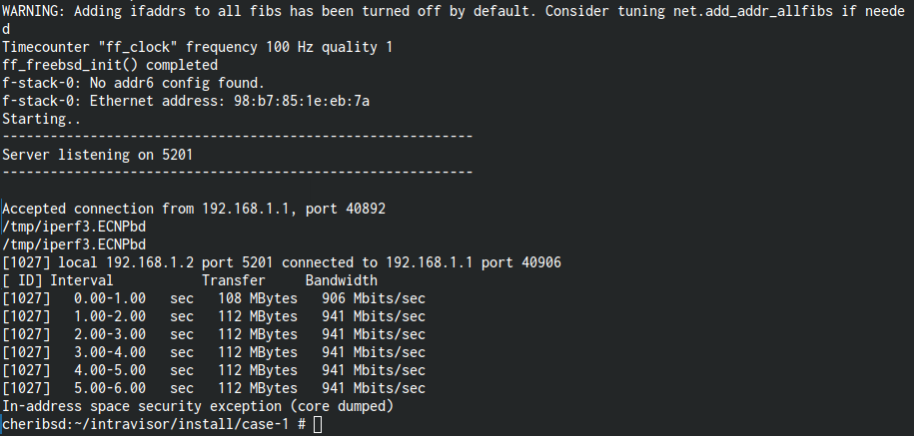}
    \caption{Applications accessing memory outside their boundaries cause exceptions under CHERI.}
    \label{fig:fault}
\end{figure}

Under CHERI, each network-stack compartment is restricted to its DDC \textit{capability}.
We verified the effectiveness of compartmentalization modifying applications to access memory ranges outside their valid boundaries.
As expected, CHERI triggers a CAP-out-of-bound exceptions, as shown in Figure~\ref{fig:fault}.

\begin{table}[b]
\centering
\caption{Results of TCP benchmarks in the three scenarios, both server and client sides. Values expressed in Mbit/s}
\label{tab:benchmark}
\begin{tabular}{|cccc|c|c|}
\hline
\multicolumn{6}{|c|}{Baseline}                                   \\ \hline
\multicolumn{1}{|c|}{Modes} & \multicolumn{2}{c|}{Server} & Efficiency & Client & Efficiency\\ \hline
\multicolumn{1}{|c|}{Baseline (\cvm{1})} & \multicolumn{2}{c|}{658} &   65.8\%    &   757 &  75.7\%    \\ \cline{1-1}
\multicolumn{1}{|c|}{Baseline (\cvm{2})} & \multicolumn{2}{c|}{658}  &  65.8\%    &   757 &  75.7\%  \\ \hline \hline

\multicolumn{6}{|c|}{Scenario 1}                                   \\ \hline
\multicolumn{1}{|c|}{Modes} & \multicolumn{2}{c|}{Server} & Efficiency & Client & Efficiency \\ \hline
\multicolumn{1}{|c|}{\cvm{1}} & \multicolumn{2}{c|}{658}   &  65.8\%     &   757  &  75.7\%  \\ \cline{1-1}
\multicolumn{1}{|c|}{\cvm{2}} & \multicolumn{2}{c|}{658}   &   65.8\%    &   757  &  75.7\% \\ \hline \hline

\multicolumn{6}{|c|}{Baseline}                                   \\ \hline
\multicolumn{1}{|c|}{Modes} & \multicolumn{2}{c|}{Server} & Efficiency & Client & Efficiency\\ \hline
\multicolumn{1}{|c|}{Baseline (\cvm{2})} & \multicolumn{2}{c|}{941}   &  94.1\%    &   941    & 94.1\% \\ \hline \hline

\multicolumn{6}{|c|}{Scenario 2 (uncontended)}                                   \\ \hline
\multicolumn{1}{|c|}{Modes} & \multicolumn{2}{c|}{Server} & Efficiency & Client  & Efficiency \\ \hline
\multicolumn{1}{|c|}{\cvm{2}} & \multicolumn{2}{c|}{941}  & 94.1\%   &  941   & 94.1\%    \\ \hline \hline

\multicolumn{6}{|c|}{Scenario 2 (contended)}                                  \\ \hline
\multicolumn{1}{|c|}{Modes} & \multicolumn{2}{c|}{Server} & Efficiency & Client & Efficiency \\ \hline
\multicolumn{1}{|c|}{\cvm{2}} & \multicolumn{2}{c|}{470}  &  94\%     &  531 &  106.2\%    \\ \cline{1-1}
\multicolumn{1}{|c|}{\cvm{3}} & \multicolumn{2}{c|}{470}  &  94\%    &  410  &  82\%  \\ \hline
\end{tabular}
\end{table}

For the network-related evaluations, we executed our modified \iperf to calculate the maximum bandwidth in both server (receiver) and client (sender) modes. We also computed the \emph{efficiency}, as the resulting bandwidth divided by the maximum bandwidth \textit{theoretically} achievable (1Gbps for each Ethernet port). 
In Scenario~1 and \textit{Baseline}, where both Ethernet ports are in use, we are not achieving high efficiency due to the hardware limitations imposed by the PCI NIC.
In all scenarios, as shown in Table~\ref{tab:benchmark}, we reached the maximum bandwidth possible with our hardware setup.
In the contented Scenario~2, we notice an unbalance among the achieved bandwidth, which can be attributed to the lack of mechanisms for fairness control. 
We defer the investigation of Quality-of-Service (QoS) approaches or the integration of DPDK QoS features to future works.

\begin{figure}[t]
    \centering
    \includegraphics[width=1\linewidth]{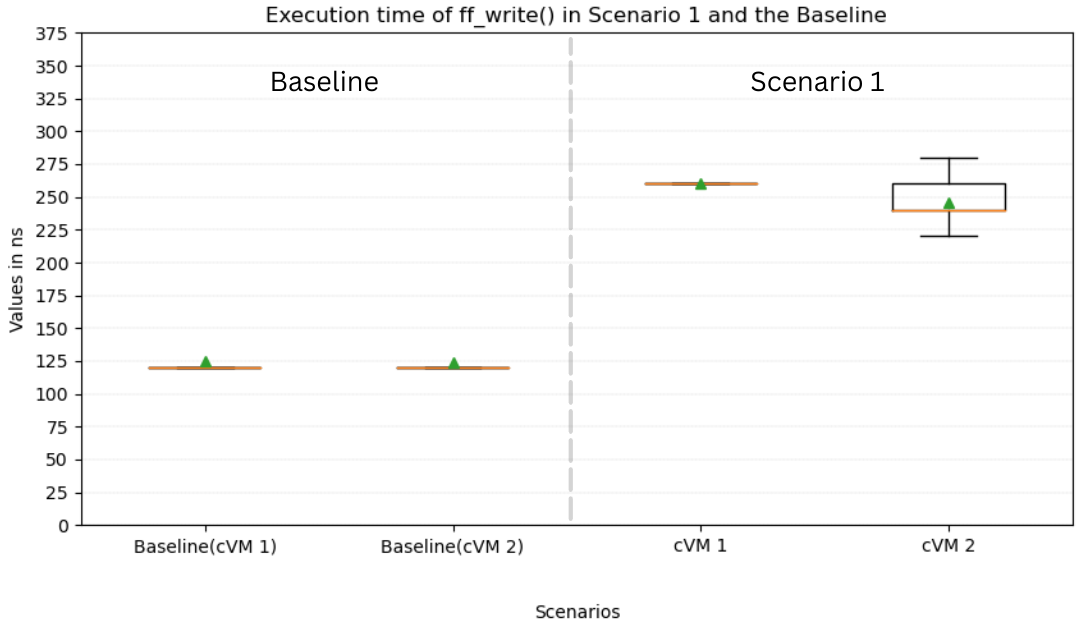}
    \caption{Execution time of the \texttt{ff\_write()} function in Scenario~1, compared with \textit{Baseline}.}
    \label{fig:ris1}
\end{figure}

\begin{figure}[t]
    \centering
    \includegraphics[width=1\linewidth]{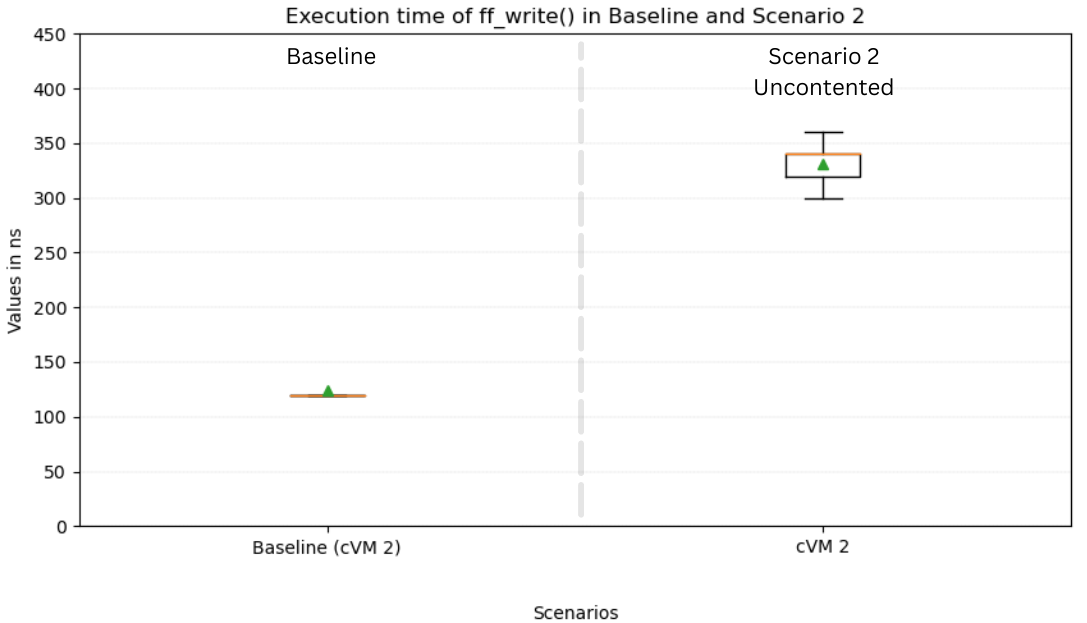}
    \caption{Execution time of the \texttt{ff\_write()} function in Scenario~2, compared with \textit{Baseline}.}
    \label{fig:ris2}
\end{figure}

\begin{figure}[t]
    \centering
    \includegraphics[width=1\linewidth]{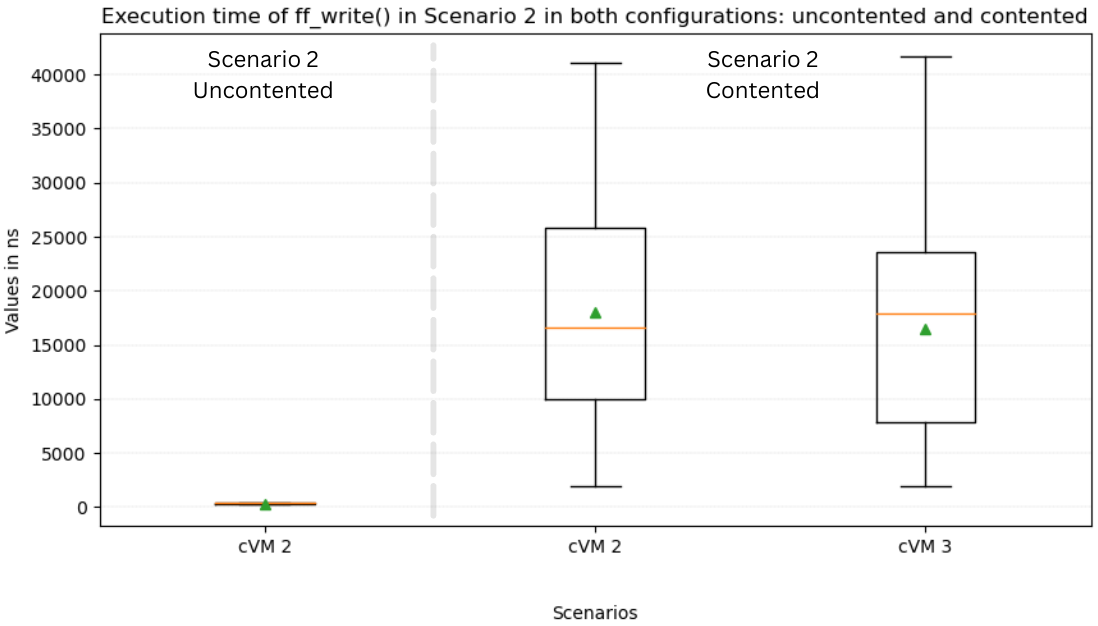}
    \caption{Execution time of the \texttt{ff\_write()} function in Scenario~2 in uncontented and contented configurations.}
    \label{fig:ris3}
\end{figure}

In the evaluation of the execution time of \texttt{ff\_write()}, each test measures the execution of 1 million iterations, and Figures~\ref{fig:ris1}-\ref{fig:ris3} present averages and standard deviations as box plots. Moreover, in all the experiments, outliers ($\approx$ 10\% of the iterations) are removed with a standard IQR strategy.
Figure~\ref{fig:ris1} compares Scenario~1 with \textit{Baseline}.
The first two boxes from the left depict the execution time for \textit{Baseline}, while the other two the execution time of Scenario~1. 
Note that, in cVMs we can't directly access the timers of the system, the execution time always includes a cross-compartment jump to the Intravisor, the execution of the syscall in CheriBSD, and the return from the Intravisor to access them. 
As seen in Figure~\ref{fig:ris1}, the impact of the CHERI compartment is minimal, and amounts to approximately 125 ns,
corresponding to the additional indirections required by the musl - Intravisor mechanism.
In both \textit{Baseline} (\cvm{1} and \cvm{2}) and \cvm{1} for Scenario~1, the 25th and 75th percentiles are identical, leading to the absence of a visible box in the box plot. This indicates that more than 50\% of the results were identical. The experiments were repeated several times, consistently yielding the same outcomes.


In Scenario~2, the cost of the cross-compartment jump and return in the cVM is included in the measurements. Moreover, due to the required synchronization between the main-loop of \mbox{F-Stack} and the execution of the \texttt{ff\_write()} (see~\ref{sec:architecture}), the measures also include the time necessary to acquire the mutex.
Figure~\ref{fig:ris2} shows the execution time measurements of the \texttt{ff\_write()} of the uncontented setup compared with \textit{Baseline}. In this case, we increased the interval between two consecutive \texttt{ff\_write()} to reduce the possibility to be blocked for a long time by the mutex.
Compared to Scenario~1, despite the additional indirection to jump between the two \cvm{s} and the handling of the mutex, the experienced slow down is contained to approximately 200 ns.

The effect of the mutex contention is evident in the contented Scenario~2, where three cVMs can acquire and wait on the mutex.
As shown in Figure~\ref{fig:ris3}, operations on the mutex result in an overhead of around 19.000 ns ($152\times$ slowdown).
Nonetheless, despite the overhead introduced for handling the mutex, the contented Scenario~2 reaches the maximum bandwidth supported by the hardware in both server (receiver) and client (sender) modes (see Table~\ref{tab:benchmark}).
As future work, we plan to investigate in details the impact of different locking strategies to further reduce the overhead of our designs.


\section{Related Work}\label{sec:relatedwork}

Several works have addressed different aspects of ensuring security in embedded systems~\cite{fut-app-iot,sec-embedd,cyb-sec-iot}. These efforts span the entire embedded system stack, from designing robust and energy-efficient hardware to developing secure, lightweight operating systems\cite{netstack} and strong encryption protocols.
For instance, the OS plays a critical role in securing UAVs, as it must implement countermeasures specific to embedded systems to defend against different security attacks~\cite{UAV-Sec}.
Prior studies on compartmentalizing network stacks in OSs, like FreeRTOS~\cite{freertos} and RIOT~\cite{riot}, have relied on coarse-grained (task-level) isolation using MPUs or MMUs, still leaving systems vulnerable to memory-related attacks such as buffer overflow~\cite{freertos-vulnerabilities, freertos-bof}. CHERI can mitigate these vulnerabilities by providing fine-grained memory access control, replacing or cooperating with MPU/MMU-based solutions~\cite{cherirtos}.
Other hardware-based solutions also address compartmentalized software design by isolating critical trusted security functions, like authentication, from the main untrusted OS. Nevertheless, they still rely on MPU-like mechanisms for isolation. For instance, Arm TrustZone~\cite{arm-trustzone} uses Secure Attribution Unit (SAU) or Implementation Defined Attribution Unit (IDAU) to separate memory into secure and insecure partitions. As a consequence, CHERI would allow enhanced internal memory protection, mitigating attacks from the untrusted OS~\cite{cheri-tree}.

CHERI-enabled architectures,
help addressing mixed-criticality environments by isolating vulnerable components.
One notable work on CHERI compartmentalization in embedded systems is CompartOS~\cite{compartOS}. The authors leverage CHERI \textit{capabilities} to isolate linkage modules by using a CHERI extension to the library \textit{libdl}.
However, as also documented in this work, this approach introduces some overhead compared to static solutions, which are typically more common in the safety-critical embedded systems.
CAP-VMs\cite{CAPVM} was also extended with a new mechanism to reduce the duplication of the libraries running in the cVMs has been implemented~\cite{orc}. This mechanism could also be used as a future integration for our work to reduce the memory utilization.

DPDK has been routinely used in all those contexts requiring low network overhead and high bandwidth (e.g.,~\cite{dpdk1,dpdk2,dpdk3}). Among those, F-Stack is the predominant TCP/IP library used, with good scalability and performance~\cite{f-stack1}.
In addition, the high-performance DPDK has been integrated into embedded systems. One example of this integration involves modifying LwIP~\cite{lwip}---a lightweight TCP/IP stack designed for embedded systems---to run on top of DPDK, exploring the use of DPDK also for constrained environments~\cite{lwip-dpdk}.

\section{Conclusion \& Future Works} \label{sec:conclusion}


This work has presented the potential of CHERI to provide strong security properties through compartmentalization of the software components of a network stack.
Specifically, we presented different trade-offs for designs that isolate DPDK, F-Stack, and a network benchmark application using Intravisor-based \cvm{s}.
Our evaluation on the CHERI-enabled Arm Morello platform has validated the effectiveness of CHERI compartmentalization and has shown the low overhead potential of our \cvm{}-based design.
However, despite achieving full bandwidth, the evaluation also underlined the impact of synchronization among different \cvm{s}.


Further research will focus on isolating all the components within the system, including finer-grained compartmentalization of the network stack and peripheral drivers. Additional scenarios include: (i)~the separation of DPDK from F-Stack and the application; and, (ii) the separation of the entire stack.
Future analysis will consider the design and potential impact of different synchronization strategies among compartments.


\begin{thebibliography}{00}
\bibitem{MxU} Runyu Pan and Gabriel Parmer. 2019. "MxU: Towards Predictable, Flexible, and Efficient Memory Access Control for the Secure IoT". ACM Trans. Embed. Comput. Syst. 18, 5s, Article 103 (October 2019), 20 pages. https://doi.org/10.1145/3358224
\bibitem{COTS} D. Dasari, B. Akesson, V. Nélis, M. A. Awan and S. M. Petters, "Identifying the sources of unpredictability in COTS-based multicore systems," 2013 8th IEEE International Symposium on Industrial Embedded Systems (SIES), Porto, Portugal, 2013, pp. 39-48, doi: 10.1109/SIES.2013.6601469.
\bibitem{dpdk-morello} spdk-morello/dpdk: Data Plane Development Kit; \url{https://github.com/spdk-morello/dpdk/tree/morello}
\bibitem{dpdk} Home - DPDK. \url{https://www.dpdk.org/}
\bibitem{CVE-2023-52370} CVE-2023-52370 | CVE. \url{https://www.cve.org/CVERecord?id=CVE-2023-52370}
\bibitem{CVE-2023-6951} CVE-2023-6951 | CVE. \url{https://www.cve.org/CVERecord?id=CVE-2023-6951}
\bibitem{CVE-2024-38951} CVE-2024-38951 | CVE. \url{https://www.cve.org/CVERecord?id=CVE-2024-38951}
\bibitem{CAPVM} Vasily A. Sartakov and Llu{\'i}s Vilanova and David Eyers and Takahiro Shinagawa and Peter Pietzuch. 2022. "CAP-VMs: Capability-Based Isolation and Sharing in the Cloud". OSDI '22. isbn: 978-1-939133-28-1. \url{https://www.usenix.org/conference/osdi22/presentation/sartakov}
\bibitem{PX4} Open Source Autopilot for Drones - PX4 Autopilot. \url{https://px4.io/}
\bibitem{PX4-paper} L. Meier, D. Honegger and M. Pollefeys, "PX4: A node-based multithreaded open source robotics framework for deeply embedded platforms," 2015 IEEE International Conference on Robotics and Automation (ICRA), Seattle, WA, USA, 2015, pp. 6235-6240, doi: 10.1109/ICRA.2015.7140074.
\bibitem{Morello} R. Grisenthwaite, G. Barnes, R. N. M. Watson, S. W. Moore, P. Sewell and J. Woodruff, "The Arm Morello Evaluation Platform—Validating CHERI-Based Security in a High-Performance System," in IEEE Micro, vol. 43, no. 3, pp. 50-57, May-June 2023, doi: 10.1109/MM.2023.3264676.
\bibitem{Sonata} Unveiling Sonata: Affordable CHERI Hardware for Embedded Systems. \url{https://www.design-reuse.com/news/55539/sonata-cheri-risc-v-prototype-board.html}
\bibitem{CheriBSD} CheriBSD. \url{https://www.cheribsd.org/}
\bibitem{CHERIISA} Robert N. M. Watson, et al., "Capability Hardware Enhanced RISC Instructions: CHERI Instruction-Set Architecture (Version 9)", Technical Report UCAM-CL-TR-987, Computer Laboratory, September 2023. \url{https://www.cl.cam.ac.uk/techreports/UCAM-CL-TR-987.pdf}
\bibitem{Mavlink} Introduction - MAVLink Devoloper Guide. \url{https://mavlink.io/en/}
\bibitem{AutoCHERI} AutoCHERI. \url{https://autocheri.tech/}
\bibitem{HWenabled} R. N. M. Watson et al. "CHERI: Hardware-Enabled C/C++ Memory Protection at Scale" in IEEE Security \& Privacy, vol. 22, no. 4, pp. 50-61, July-Aug. 2024, doi: 10.1109/MSEC.2024.3396701.
\bibitem{UAV-Sec} S. Iqbal, "A Study on UAV Operating System Security and Future Research Challenges," 2021 IEEE 11th Annual Computing and Communication Workshop and Conference (CCWC), Las Vegas, NV, USA, 2021, pp. 0759-0765, doi: 10.1109/CCWC51732.2021.9376151.
\bibitem{lwip-dpdk}R. Rajesh, K. B. Ramia and M. Kulkarni, "Integration of LwIP Stack over Intel(R) DPDK for High Throughput Packet Delivery to Applications," 2014 Fifth International Symposium on Electronic System Design, Surathkal, India, 2014, pp. 130-134, doi: 10.1109/ISED.2014.34.
\bibitem{compartOS}Hesham Almatary et al. "CompartOS: CHERI Compartmentalization for Embedded Systems," 2022.\url{https://arxiv.org/abs/2206.02852}
\bibitem{fut-app-iot} H. U. Rehman, M. Asif and M. Ahmad, "Future applications and research challenges of IOT," 2017 International Conference on Information and Communication Technologies (ICICT), Karachi, Pakistan, 2017, pp. 68-74, doi: 10.1109/ICICT.2017.8320166.
\bibitem{sec-embedd}Srivaths Ravi, Anand Raghunathan, Paul Kocher, and Sunil Hattangady. 2004. "Security in embedded systems: Design challenges. ACM Trans. Embed. Comput. Syst. 3, 3 (August 2004), 461–491. https://doi.org/10.1145/1015047.1015049
\bibitem{cyb-sec-iot}A. Aldahmani, B. Ouni, T. Lestable and M. Debbah, "Cyber-Security of Embedded IoTs in Smart Homes: Challenges, Requirements, Countermeasures, and Trends," in IEEE Open Journal of Vehicular Technology, vol. 4, pp. 281-292, 2023, doi: 10.1109/OJVT.2023.3234069.
\bibitem{fstack} F-Stack | High Performance Network Framework Based On DPDK. \url{https://www.f-stack.org/}
\bibitem{emb-sys-sec} D. Papp, Z. Ma and L. Buttyan, "Embedded systems security: Threats, vulnerabilities, and attack taxonomy," 2015 13th Annual Conference on Privacy, Security and Trust (PST), Izmir, Turkey, 2015, pp. 145-152, doi: 10.1109/PST.2015.7232966.
\bibitem{sec-iiot}A. C. Panchal, V. M. Khadse and P. N. Mahalle, "Security Issues in IIoT: A Comprehensive Survey of Attacks on IIoT and Its Countermeasures," 2018 IEEE Global Conference on Wireless Computing and Networking (GCWCN), Lonavala, India, 2018, pp. 124-130, doi: 10.1109/GCWCN.2018.8668630.
\bibitem{trends-rahimi}P. Rahimi, A. K. Singh, X. Wang and A. Prakash, "Trends and Challenges in Ensuring Security for Low-Power and High-Performance Embedded SoCs," 2021 IEEE 14th International Symposium on Embedded Multicore/Many-core Systems-on-Chip (MCSoC), Singapore, Singapore, 2021, pp. 226-233, doi: 10.1109/MCSoC51149.2021.00041.
\bibitem{musl} musl libc. \url{https://musl.libc.org/}
\bibitem{lkl} O. Purdila, L. A. Grijincu and N. Tapus, "LKL: The Linux kernel library," 9th RoEduNet IEEE International Conference, Sibiu, Romania, 2010, pp. 328-333.
\bibitem{umtx} \_umtx\_op(2) - FreeBSD Manual. \url{https://man.freebsd.org/cgi/man.cgi?query=_umtx_op&sektion=2&n=1}
\bibitem{iperf3} iPerf - The TCP, UDP and SCTP network bandwidth measurement tool. \url{https://iperf.fr/}
\bibitem{f-stack1} A. B. Narappa, F. Parola, S. Qi and K. K. Ramakrishnan, "Z-Stack: A High-Performance DPDK-Based Zero-Copy TCP/IP Protocol Stack," 2024 IEEE 30th International Symposium on Local and Metropolitan Area Networks (LANMAN), Boston, MA, USA, 2024, pp. 100-105, doi: 10.1109/LANMAN61958.2024.10621881.
\bibitem{dpdk1}Abhishek Bhattacharyya, Shunmugapriya Ramanathan, Andrea Fumagalli, Koteswararao Kondepu. An end-to-end DPDK-integrated open-source 5G standalone Radio Access Network: A proof of concept. Computer Networks. \url{https://doi.org/10.1016/j.comnet.2024.110533}.
\bibitem{dpdk2}V. Bode, C. Trinitis, M. Schulz, D. Buettner and T. Preclik, "Adopting User-Space Networking for DDS Message-Oriented Middleware," 2024 IEEE International Conference on Pervasive Computing and Communications (PerCom), Biarritz, France, 2024, pp. 36-46, doi: 10.1109/PerCom59722.2024.10494460.
\bibitem{dpdk3}J. Umeike, S. Agarwal, N. Lazarev and M. Alian, "Userspace Networking in gem5," 2024 IEEE International Symposium on Performance Analysis of Systems and Software (ISPASS), Indianapolis, IN, USA, 2024, pp. 179-191, doi: 10.1109/ISPASS61541.2024.00026.
\bibitem{orc} Vasily A. Sartakov, Lluis Vilanova, Munir Geden, David Eyers, Takahiro Shinagawa, Peter Pietzuch. "ORC: Increasing Cloud Memory Density via Object Reuse with Capabilities". In 17th USENIX Symposium on Operating Systems Design and Implementation (OSDI 23) (pp. 573-587). USENIX Association.
\bibitem{netstack} Zuepke, Alexander, et al. "For Safety and Security Reasons: The Cost of Component-Isolation in IoT" Logistics, Supply Chain, Sustainability and Global Challenges, vol. 7, no. 1, Sciendo, 2016, pp. 41-50. https://doi.org/10.1515/jlst-2016-0004
\bibitem{lwip} LwIP - A Lightweight TCP/IP stack - \url{https://savannah.nongnu.org/projects/lwip/}
\bibitem{mavlink} Introduction - MAVLink Developer Guide - \url{https://mavlink.io/en/}
\bibitem{freertos-vulnerabilities} Al-Boghdady, Abdullah, Khaled Wassif, and Mohammad El-Ramly. 2021. "The Presence, Trends, and Causes of Security Vulnerabilities in Operating Systems of IoT’s Low-End Devices" Sensors 21, no. 7: 2329. https://doi.org/10.3390/s21072329
\bibitem{freertos-bof} G. Mullen and L. Meany, "Assessment of Buffer Overflow Based Attacks On an IoT Operating System," 2019 Global IoT Summit (GIoTS), Aarhus, Denmark, 2019, pp. 1-6, doi: 10.1109/GIOTS.2019.8766434.
\bibitem{cherirtos} H. Xia et al., "CheriRTOS: A Capability Model for Embedded Devices," 2018 IEEE 36th International Conference on Computer Design (ICCD), Orlando, FL, USA, 2018, pp. 92-99, doi: 10.1109/ICCD.2018.00023.
\bibitem{cheriot-netstack} S. Amar et al. "CHERIoT: Complete Memory Safety for Embedded Devices", 2023, In Proceedings of the 56th Annual IEEE/ACM International Symposium on Microarchitecture (MICRO '23). Association for Computing Machinery, New York, NY, USA, 641–653. https://doi.org/10.1145/3613424.3614266
\bibitem{cheri-tree} T. Van Strydonck et al., "CHERI-TrEE: Flexible enclaves on capability machines," 2023 IEEE 8th European Symposium on Security and Privacy (EuroS\&P), Delft, Netherlands, 2023, pp. 1143-1159, doi: 10.1109/EuroSP57164.2023.00070.
\bibitem{arm-trustzone} A. Arm, “Arm TrustZone Technology for the Armv8-M Architecture” ARM Limited, 2018, \url{https://developer.arm.com/documentation/100690/}.
\bibitem{riot} E. Baccelli et al., "RIOT: An Open Source Operating System for Low-End Embedded Devices in the IoT," in IEEE Internet of Things Journal, vol. 5, no. 6, pp. 4428-4440, Dec. 2018, doi: 10.1109/JIOT.2018.2815038
\bibitem{freertos} Fei Guan, Long Peng, Luc Perneel, and Martin Timmerman. 2016. Open source FreeRTOS as a case study in real-time operating system evolution. J. Syst. Softw. 118, C (August 2016), 19–35. \url{https://doi.org/10.1016/j.jss.2016.04.063}
\end{thebibliography}
\end{document}